\documentclass[12pt,preprint]{aastex}

\begin{document}

\title{The local power spectrum and correlation hierarchy of the cosmic
mass field}

\author{Hu Zhan, Priya Jamkhedkar, and Li-Zhi Fang}

\affil{Department of Physics, University of Arizona, Tucson,
AZ 85721}

\begin{abstract}
We analyze the power spectrum of a QSO's Ly$\alpha$ transmitted flux in the
discrete wavelet transform (DWT) representation. Although the mean DWT power
spectrum is consistent with its counterpart in Fourier representation,
the spatial distribution of the local power varies greatly, i.e. the
local DWT power spectra show remarkably spiky structures on small scales.
To measure these spiky features, we introduce the quantities {\it roughness}
of the local power spectrum, and the {\it correlation} between spikes on
different scales. We then test the predictions made by the correlation 
hierarchy model on the roughness and the scale-scale correlations of 
the local power spectrum. Using the Ly$\alpha$ transmitted flux of the 
QSO HS1700, we find that the underlying cosmic mass field of the 
transmitted flux at redshift around $z \simeq 2.2$ can be described by 
the hierarchical clustering model on physical scales from 2.5 h$^{-1}$ Mpc 
to few tens h$^{-1}$ kpc in an Einstein-de Sitter universe. However, 
the non-linear 
features of the clustering show differences on different scale ranges; 1. 
On physical scales larger than $\sim 1.3$ h$^{-1}$ Mpc, the field is 
almost Gaussian. 2. On scales 1.3 h$^{-1}$ Mpc - $0.3$ h$^{-1}$ Mpc, 
the field is consistent with the correlation hierarchy 
with a constant value for the coefficient $Q_4$. 3. On scales less than 
300 h$^{-1}$ kpc, the field is no longer Gaussian, but
essentially intermittent. In this case, the field can still be fitted by 
the correlation 
hierarchy, but the coefficient, $Q_4$, should be scale-dependent. These 
three points are strongly supported by the following result: the scale 
dependencies of $Q_4$ given by two statistically independent measures, 
i.e. $Q_4^R$ by the roughness and $Q_4^C$ by scale-scale correlation, 
are the same in the entire scale range considered.

\end{abstract}


\section{Introduction}

The most popular statistical measure of the cosmic mass density
field $\rho({\bf x})$ is the Fourier power spectrum. For a
homogeneous and isotropic random mass field, the Fourier power
spectrum $P(k)$ is given by
%eq1
\begin{equation}
\langle \delta_{\bf k}\delta_{\bf k'} \rangle =
  P(k)\delta^K_{{\bf k,k'}},
\end{equation}
where $\langle ...\rangle$ is the average over an ensemble,
$\delta_{\bf k}$ is the Fourier transform of the density contrast
$\delta ({\bf x}) = (\rho({\bf x})-\bar{\rho})/\bar{\rho}$,
$\bar{\rho}$ is the average density and $\delta^K_{{\bf k,k'}}$ the
Kronecker-Delta function. Although the power spectrum is only a
second order statistical measure of the
inhomogeneity of the random density field, it directly reflects the
scales on which non-linear physical processes affect structure
formation.

The Fourier power spectrum, however, loses spatial information completely
because of the non-local nature of the Fourier mode. Thus, it cannot
be used to describe position-related statistical features of the mass
field. In other words, the power spectrum, or the galaxy-galaxy
correlation function cannot detect large scale filaments and sheets
in the galaxy distribution. This disadvantage is clearly seen when
reconciling the power spectrum description with the singular
behavior of the cosmic mass field (Navarro, Frenk, \& White
1997; Moore et al. 2000; Jing \& Suto 2000). The existence of singular
structures, like massive halos with density profiles
$\rho (r)\propto r^{-\alpha}$ ($\alpha >0$) indicates that the power
of mass density perturbations on scale $r \propto 1/k$ is not
uniformly distributed in space, but
concentrated in rare high power regions. This problem calls for a
description of {\it local} power spectrum, which provides the
information of power distribution with respect to both the scale and
physical position of the fluctuation.

The local power spectrum of the underlying mass field of a QSO's Ly$\alpha$
forest has been recently analyzed (Jamkhedkar, Zhan, \& Fang 2000). It
is found that the local power spectrum of the transmitted flux of the QSO's
absorption shows prominent spiky structures on small scales. That is,
the transmitted flux consists of rare but strong density fluctuations
randomly scattered in space with very low power of fluctuations in
between. Moreover, the spiky features are more significant on
smaller scales. This indicates an excess of large fluctuations on small 
scales in comparison to a Gaussian distribution, i.e. the random mass field 
traced by the Ly$\alpha$ transmitted flux probably is intermittent
(Zeldovich, Ruzmaikin, \& Sokoloff 1990; Frisch 1995; Shraiman \& Siggia 
2000). The local power spectrum is especially useful for describing the
non-linear features of the cosmic mass field, and thus for testing models
of the non-linear clustering of the cosmic mass field.

The purpose of this paper is to test the most popular non-linear clustering
model -- the hierarchical clustering model. The hierarchical clustering
scenario (e.g. White 1979) assumes that the non-linear cosmic mass field
satisfies a linked-pair approximation, or correlation hierarchy, i.e.
$\xi_n\simeq Q_n\xi_2^{n-1}$, where $\xi_n$ and $\xi_2$
are the $n$- and 2-point correlation functions of mass density,
respectively. The coefficients $Q_n$s are assumed to be constant or
scale-independent. This model has been widely applied to construct
semi-analytic models of gravitational clustering in the universe.
With the correlation hierarchy, all high order correlation functions are
given by two-point correlation functions and coefficients $Q_n$s. Therefore,
in hierarchical clustering model, the spiky features of the local power 
spectrum should also be produced by the two point correlation function
and $Q_n$s. Thus, the applicable range of the correlation hierarchy
can be determined by comparing these predictions with observed
local power spectrum.

The paper is organized as follows. The first half (\S 2 and \S 3) 
studies the statistical features of the local power spectrum of the
underlying mass field revealed by the QSO's Ly$\alpha$ forest. The second
half (\S 4 and \S 5) investigates the local power spectrum of the
hierarchically clustered field. \S 2 provides a brief background of local
power spectrum in the discrete wavelet transform (DWT) representation. The
local power spectra measured from Ly$\alpha$ transmitted flux of QSO
HS1700+64 are presented in \S 3. The focus there is on the spiky features
and roughness of the local power spectrum. In \S 4, we investigate the
predication of roughness and scale-scale correlation of the spiky structures 
given by the linked-pair approximation. We then test these predictions with
the observed local power spectrum, and obtain the applicable range of the
correlation hierarchy for the cases of a constant and scale-dependent 
$Q$. \S 5 consists of the conclusions and discussions.

\section{Local power spectrum}

\subsection{Power spectrum in the DWT representation}

For the sake of simplicity, we analyze a 1-D density distribution
sample $\rho(x)$ in the range $0<x<L$, which is assumed to be a
stationary random field. It is straightforward to extend the results to
2-D and 3-D.

The first generation of the algorithm of local power spectrum is
the windowed Fourier analysis (Gabor 1946), which decomposes a
distribution $\rho(x)$ as
%eq2
\begin{equation}
\hat{\rho}(k,x_0) = \int \rho(x)g(x - x_0)e^{-ikx} dx,
\end{equation}
where $g(x - x_0)$ is a window function of size $\Delta x$ around
position $x_0$. We thus define a local power spectrum at $x_0$ as
$|\hat{\rho}(k,x_0)|^2$. However, for a given spatial size $\Delta x$,
the uncertainty in the wavenumber is $\Delta k \simeq 2\pi/\Delta x$.
Consequently, the local power spectrum $|\hat{\rho}(k,x_0)|^2$ is
uncertain on scales $ k \leq  2\pi/\Delta x$. This problem leads
to the development of the discrete wavelet transform (DWT), which
decomposes $\rho(x)$ with orthogonal and complete bases on successive
scales obeying the condition $\Delta k \Delta x\simeq 2\pi$.
Indeed, the transform in eq.(2) is a predecessor of the DWT.

To apply the DWT, we first chop $L$ into $2^j$ subintervals,
each of which spans a spatial range $L/2^j$ labeled with $l=0, ...2^j-1$,
and the subinterval $l$ is from $Ll/2^j$ to $L(l+1)/2^j$.
The density contrast, $\delta(x)=(\rho(x)-\bar{\rho})/\bar{\rho}$, can be
decomposed as\footnote{To be exact, the distribution $\delta(x)$
in eqs.(3) and (4) is a periodic extension of the density field
over an interval of length $L$ (Fang \& Feng 2000).
}
%eq3
\begin{equation}
\delta(x) = \sum_{j=0}^{\infty} \sum_{l=0}^{2^j-1}
  \tilde{\epsilon}_{j,l} \psi_{j,l}(x),
\end{equation}
where $\psi_{j,l}(x)$ with $j= 0, 1,...$ and $l=0,...2^j-1$ are the
orthogonal and complete bases of the discrete wavelet transform (DWT)
(Daubechies 1992; Fang \& Thews 1998). The non-zero range of
$\psi_{j,l}(x)$ is mainly between $lL/2^j$ and $(l+1)L/2^j$. The wavelet
function coefficient (WFC), $\tilde{\epsilon}_{j,l}$, in eq.(3) is
obtained by projecting $\delta(x)$ onto $\psi_{j,l}(x)$
%eq4
\begin{equation}
\tilde{\epsilon}_{j,l}=
\int \delta(x) \psi_{j,l}(x)dx.
\end{equation}
The $\tilde{\epsilon}_{j,l}$ describes the density perturbation at the
position $lL/2^j$ on the length scale $L/2^j$ (or, the wavelet scale $j$).
For a Haar wavelet the WFC,
$\tilde{\epsilon}_{j,l}$, is the difference between the mean density
contrasts in ranges $lL/2^j \leq x < (l+1/2)L/2^j$ and
$(l+1/2)L/2^j \leq x < (l+1)L/2^j$.  For other wavelets, the WFC,
$\tilde{\epsilon}_{j,l}$, is also a measure of the density contrast
difference
on a scale $L/2^j$ at a position $l$. We will use the Daubechies 4 wavelet
(Daubechies 1992) in our numerical calculation below.

The decomposition eq.(3) preserves all the information contained in the
original field. Consider a sample with a resolution of $L/2^J$, which is
equivalent to $2^J$ grid-points on $L$. The degree of freedom of the
sample is then $2^J-1$, where the condition of average $\bar{\delta}=0$
reduces one degree of freedom. Since the local density fluctuations
have to be measured over at least two neighboring grids, the smallest length,
on which the WFCs can be calculated, is
$2\times (L/2^J)$. The corresponding scale, $j$, is then $J -1$. Thus, the
total number of the WFCs ($j=1...J-1$, and $l=0...2^j$) is
%eq5
\begin{equation}
\sum_{j=1}^{J-1} 2^{J-j} = 2^J-1.
\end{equation}
Therefore, the WFCs contain complete information of the random field
in the DWT representation. The original distribution, $\delta(x)$,
can be exactly and unredundantly reconstructed from the WFCs.

Parseval's theorem for the DWT decomposition is (Fang \& Thews
1998)
%eq6
\begin{equation}
\frac{1}{L}\int_0^L |\delta(x)|^2 dx = \sum_{j= 0}^{\infty}
\frac{1}{L}\sum_{l=0}^{2^j-1} \tilde{\epsilon}_{j,l}^2,
\end{equation}
which implies that the power of perturbations can be decomposed into
modes $(j,l)$. The power of the mode $(j,l)$ is given by
$\tilde{\epsilon}_{j,l}^2$. Thus the power spectrum in the DWT
representation is given by $\langle\tilde{\epsilon}_{j,l}^2\rangle$, where
$\langle ...\rangle$ is for the ensemble average.
If the random field is stationary (or homogeneous in higher dimensions),
$\langle\tilde{\epsilon}_{j,l}^2\rangle$ has to be independent
of $l$. One can then define the DWT power spectrum as
%eq7
\begin{equation}
P_j =\langle\tilde{\epsilon}_{j,l}^2\rangle.
\end{equation}
If the ``fair sample hypothesis" (Peebles 1980) holds, instead of
the ensemble average of eq.(6), we can use the average over $l$, i.e.
%eq8
\begin{equation}
P_j =\frac{1}{2^j}\sum_{l=0}^{2^j-1}
 \tilde{\epsilon}_{j,l}^2
\end{equation}
It has been shown that $P_j$ is actually a band-averaged Fourier
power spectrum $P(k)$ around $k = 2\pi 2^j/L$ (Pando \& Fang 1998;
Fang \& Feng 2000), i.e.
%eq9
\begin{equation}
P_j = \frac{1}{2^j} \sum_{n = - \infty}^{\infty}
 |\hat{\psi}(n/2^j)|^2 P(k),
\end{equation}
where $P(k)$ is the Fourier power spectrum with the wavenumber
$k=2\pi n/L$, and $\hat{\psi}(n/2^j)$ is the Fourier transform of
the basic wavelet $\psi(x)$. The non-zero range of $\hat{\psi}(n/2^j)$
is only around $n/2^j \simeq n_g$, where $n_g$ depends on wavelets.
For the Daubechies 4 wavelet, $n_g\simeq \pm 1$.

The wavenumber difference between scales $j$ and $j+1$ is
$\Delta k= 2\pi(2^{j+1} -2^j)/L =2\pi/\Delta x$, with
$\Delta x =L/2^j$. Therefore, the resolution of the DWT power spectrum
is not as dense as the Fourier power spectrum. The Fourier modes are
uniformly distributed over the scale space, while the $P_j$'s are
distributed
on octaves. However, this does not mean that the DWT decomposition
losses information (Fang \& Feng 2000). Besides the DWT power spectrum
$P_j$, the second order DWT statistics also provide information about the
spatial distribution of the power. Thus, an advantage of the DWT
representation is its ability to calculate the local power spectrum.

\subsection{The local DWT power spectrum}

The Fourier power spectrum $P(k)$ lacks phase information, and,
therefore, cannot reveal the position-related features of clustering.
The DWT power spectrum $P_j$ doesn't either, but the DWT mode $(j,l)$
contains information of positions as it is localized at $l$. Therefore, the
power of density fluctuations at the position $l$ and on the scale $j$
is measured by
%eq10
\begin{equation}
P_{j,l}=\tilde{\epsilon}_{j,l}^2.
\end{equation}
That is, $P_{j,l}$ vs. $j$ is the power spectrum of density
perturbations localized at the position $l$.

We can also define the band-averaged local DWT power spectrum. First, 
we chop $L$
into $2^{j_s}$ ($j_s \leq j$) sub-intervals labeled as
$l_s= 0, 1...(2^{j_s}-1)$. Each sub-interval has a length $L/2^{j_s}$.
Thus, the local DWT power spectrum averaged in the sub-interval $l_s$
is given by
%eq11
\begin{equation}
P_{j,\{j_s,l_s\}}=\frac{1}{2^{j-j_s}}
 \sum_{l=l_s2^{j-j_{s}}}^{(l_s+1)2^{j-j_{s}}-1}P_{j,l}
=
\frac{1}{2^{j-j_s}}
 \sum_{l=l_s2^{j-j_{s}}}^{(l_s+1)2^{j-j_{s}}-1}
 |\tilde{\epsilon}_{j,l}|^2.
\end{equation}
$P_{j,\{j_s,l_s\}}$ generally varies with $l_s$, and it measures the spatial
distribution of the power on scale $j$ among the sub-intervals. Therefore,
the spatial distribution $P_{j,\{j_s,l_s\}}$ actually is the
spatial distribution of $P_{j,l}$ smoothed on the scale $j_s$.

Obviously, when $j_s=j$ (and so $l=l_s$),  the local power spectrum eq.(11)
reduces to eq.(10), i.e.
%eq12
\begin{equation}
P_{j,\{j,l \}}=P_{j,l}
\end{equation}
On the other hand, when $j_s=0$ (and so $l_s=0$), i.e. the smoothing is
done over the entire length $L$, $P_{j,\{j_s,l_s\}}$ is the power
averaged over the entire region $L$, i.e.
%eq13
\begin{equation}
P_{j,\{0,0\}} = P_j= \frac{1}{2^j}\sum_{l=0}^{2^{j}-1}
 \tilde{\epsilon}_{j,l}^2.
\end{equation}
Therefore, $P_{j,\{j_s,l_s\}}$ provides a multiresolutional view of the
phase space distribution of the power of the
density perturbations.

\section{Local power spectra of Ly$\alpha$ transmitted flux}

\subsection{Samples and power spectra}

High resolution spectra of Ly$\alpha$ forest QSOs are good candidates
to study the local power spectrum of the cosmic mass
field. The transmission $F(x)$ of Ly$\alpha$ forests is due to the 
absorption by gases in cool and low density regions. The pressure 
gradients are generally smaller than gravitational forces. The 
distribution of cool baryonic diffuse matter is almost point-by-point 
proportional to the underlying dark matter density (Bi, Ge \& Fang 1995). 

We use the normalized Ly$\alpha$ transmitted flux of QSO HS1700+64 
($z$ = 2.72) for
our analysis. This sample has been employed to study the evolution of
structures (Bi \& Davidsen 1997), and the Fourier and DWT power spectra
(Feng \& Fang 2000). The data ranges from 3727.012\AA $ $ to 5523.554\AA, 
with a total of 55882 pixels. In this paper, we use the data from 
$\lambda =$ 3815.6\AA $ $ to 4434.3\AA, which correspond to 
$z = 2.14 \sim 2.65$. The lower limit of the wavelength is set to 
exclude Ly$\beta$ absorption. On
average, a pixel is about 0.028\AA, equivalent to a proper distance of $\sim 5$
h$^{-1}$ kpc at $z \sim 2$ for an Einstein-de Sitter universe. Moreover, 
we subject
DWT directly to pixels without transforming them to physical positions.
The physical position is related to the pixel number in a non-linear way,
but the departure from linearity is very small and it changes very smoothly
across the sample. Since we are interested in the statistical properties of
the mass density field on scales much smaller than the whole range of the
data, we can ignore the effect of such non-linear relation in our present
analysis.

A possible source of contamination comes from the presence of metal
lines. We tried three ways to estimate the error. One way is to block the
significant metal line regions identified by Dobrzycki and Bechtold
(1996), and Scott\footnote{See 
http://qso.as.arizona.edu/ $\tilde{ }$ jscott/Spectra/index.html}.
Since the WFC $\tilde{\epsilon}_{j,l}$ is
localized, the metal line regions have been separated from the rest.
The metal line effect can be removed by not counting the DWT modes in
the blocked regions. The second way is to fill those
regions with random data which has the same mean power as the rest of
the original data, and to smooth the data over the boundaries. The third
way is to discard the metal line chunks and smoothly connect the
good chunks of data. The justification of doing this comes from
the fact that we are interested only in the statistical properties
of the density field and not in the exact location of a spike in space.
We found that different methods of removing metal lines yield different
details in the local power spectrum. However, the statistical measures of
the local power spectrum are not sensitive to the method of removing metal
lines. That is, the uncertainty of metal lines is under control.

Another source of contamination is the noise. To estimate the effect of the noise
we smooth the QSO's spectrum by filtering out all extremely sharp spikes
in the local power spectra on finest scales, which are caused by relatively
strong fluctuations between two neighboring pixels. Since such events are on
the smallest scales only, the analysis on larger scales actually does not
depend on whether we smooth the sample or not.

For this sample, the DWT power spectrum $P^F_j$ of the transmitted flux is 
shown in Fig. 1, where $P^F_j$ is defined in the same way as eq.(13), 
but the WFCs are calculated by the transmission $F(x)$, i.e. 
$\tilde{\epsilon}^F_{j,l}=\int F(x)\psi_{j,l}(x)dx$. 
The power spectrum of the metal-line-removed-and-smoothed (MLRAS) data, 
is also shown in Fig. 1. Corresponding to the DWT scale, $j$, the length scale
equals $2^{15-j}\times 5 h^{-1}$ kpc. The MLRAS power spectrum is
the same as the original data in the range $6 \leq j \leq 11$. The 
smoothing reduces the power on small scales ($j>11$). However, the 
differences between the original and MLRAS power spectra, whether 
on small scales or large scales, are much lower than the variance, hence 
the differences are statistically insignificant.

\subsection{Spiky structures of local power spectrum}

Fig. 2 shows the normalized local power spectra $P^F_{j,l}$/$P^F_j$. Fig. 3
is $P^F_{j,\{j_s,l_s\}}$/$P^F_j$ with $j_s=8$, i.e. the local power
spectrum averaged on a scale of $0.64$ h$^{-1}$ Mpc. The right panels of
Figs. 2 and 3 represent the corresponding local power spectra of the
phase-randomized (PR)
data, which is obtained by taking the inverse transform of the Fourier
coefficients of the original data after randomizing their phases uniformly
over $[0,2\pi]$ without changing the amplitudes. Therefore,
the mean powers of the left and the right panels of Figs. 2 and 3 are
actually the same on the same scale.

Two main features can be observed in Fig. 2 and 3:
First, the local power spectra of the real data (left panels) are
significantly different from their counterparts of the phase randomized
sample (right panels). The former show spiky structures, while the later
are typical noisy distributions. The
spiky structures show that the power of the flux  perturbations is
concentrated within high spikes only, while the power between two
spikes is very low, or practically zero. The spiky structures are
still remarkable on all scales $j \geq 10 $ even in the local power
spectrum averaged on scale $j_s=8$. 

Second, the spiky structures are more significant on smaller scales, or
larger $j$. That is, the ratio between the amplitudes of the spikes and
the mean power is higher for smaller scales. In other words, the
probability distribution function (PDF) of WFCs $\tilde{\epsilon}_{j,l}$ is
significantly long-tailed on small scales. This is a typical behavior
of an intermittent field (Shraiman \& Siggia 2000.)

It is well known that the random velocity field of baryonic matter generally
reduces the power of density perturbations on scales equal or less than the
velocity dispersion. As a consequence, the powers (Fig. 1) on small
scales generally must be lower than the powers of the underlying mass field
of the Ly$\alpha$ absorbers. However, Fig. 2 and 3 show that neither the
spiky feature of $P^F_{j,l}$ nor that of $P^F_{j,\{ 8,l_s\}}$ on small scales is
seriously affected by the random velocity field. This is because the DWT modes are
localized in $j$ and $l$, a spike on local mode $(j,l)$ can only be affected
by a large velocity fluctuation on the same mode $(j,l)$, regardless the
velocity fluctuation on other modes. This reduces the effect of
the velocity
field on the spiky structures.

It should be pointed out that the spikes are not always located at the
peaks of mass density field, or the position of low transmitted flux. The 
density peaks are the areas with density $\rho(x) \gg \bar{\rho}$ or the 
density contrast $\delta(x) \gg 1$,
while the spikes of the WFCs $\tilde{\epsilon}^F_{j,l}$ correspond to a high
{\it difference} between densities $|\rho(x_1)-\rho(x_2)|$ or flux 
$|F(x_1)-F(x_2)|$ on scales $|x_1-x_2| \simeq L/2^j$.
It is not necessary that a high-density difference (or high transmission 
difference) event be always located in high density halos (or low
 transmission region). For instance, Fig. 1 shows that the mean power of 
transmission fluctuations
at $j=13$ is as small as $P^F_{13}= 2\times10^{-4}$. Thus, even a spike 
$>10\sigma$ at $j=13$ means only $|F(x+L/2^{13}) -F(x)| \simeq 0.04 \bar{F}$,
which can occur at positions with either high or low $F(x+L/2^{13})$ and 
$F(x)$. Therefore, the local power spectrum of the cosmic mass field can be 
studied with samples not necessarily in high density clumps but also in 
low density regions like QSOs' Ly$\alpha$ absorption.

\subsection{Roughness of the local power spectrum}

The spiky structures, or the spatial fluctuations of the clustering power,
can be measured by the standard deviation of the distribution of the local
power $P^F_{j,l}$. It is given by
%eq14
\begin{equation}
\sigma^{p}_j=
  \left[ \langle (P^F_{j,l})^2 \rangle -(P^F_j)^2\right]^{1/2}
   = \left [\frac{1}{2^j}
     \sum_{l=0}^{2^j-1}(P^F_{j,l})^2 -(P^F_j)^2\right]^{1/2}.
\end{equation}
Obviously, the ratio $\sigma^{p}_{j}/P^F_j$ is larger if spiky
structure is stronger. Fig. 1 also plots $P^F_j+\sigma^{p}_{j}$.
It shows that the spatial fluctuations of $P^F_{j,l}$ are stronger on
smaller scales (larger $j$).

It is more convenient to measure the spiky structures by the
{\it roughness} of the local power spectrum, defined as
%eq15
\begin{equation}
R^F_j \equiv \frac{\langle (\tilde{\epsilon}^F_{j,l})^4\rangle}
    {3\langle (\tilde{\epsilon}^F_{j,l})^2 \rangle^2} -1
=\frac{(1/2^j)\sum_{l=0}^{2^j-1}(\tilde{\epsilon}^F_{j,l})^4}
 {3[(1/2^j)\sum_{l=0}^{2^j-1}(\tilde{\epsilon}^F_{j,l})^2]^2}-1 .
\end{equation}
The roughness is essentially the same as eq.(14), i.e., given by the 4-th
order moment of the WFCs. The definition of eq.(15) includes only
the irreducible correlation. It ensures that $R_j=0$ for
a Gaussian PDF of $\tilde{\epsilon}_{j,l}$, and always $R_j>-1$, i.e. 
a non-zero $R_j$ is from non-Gaussian clustering.   

Fig. 4 presents the roughness of the sample HS1700+64. It shows a
significant increase in the roughness on scales less than 0.3 h$^{-1}$ Mpc. 
The roughness of the MLRAS sample is also plotted in Fig. 4. The MLRAS 
sample yields about the same roughness as the original data. The difference 
on scale $j\geq12$ is due to the smoothing.

\subsection{Power spectrum and roughness of the mass field}

Previous sections give the DWT power spectrum $P_j^F$ and roughness
$R^F_j$ for the transmission. To compare with linked 
pair model (\S 4), we need the DWT power spectrum $P_j$ and roughness 
$R_j$ of the underlying mass field traced by Ly$\alpha$ forests. The 
relation between the $P_j$ and $P_j^F$ or $R_j$ and $R^F_j$ is not 
trivial, because the relationship between the Ly$\alpha$ optical 
depth $\tau$ and mass density contrast $\delta$ of baryonic matter 
$\rho_b$ is nonlinear, i.e. 
$\tau = A(\rho_b/\bar{\rho}_b)^a=A(1+\delta)^a$ with $a= 1.5 - 1.9$.
(e.g. Hui \& Gnedin 1997). However, what we need actually is only the 
shape of the DWT power spectrum of mass field. The over-all normalization 
of the power spectrum is not important. Therefore, we will only try to show 
that the shape of $P_j$ can be effectively constrained by the  Ly$\alpha$
transmission. 

The WFCs, $\tilde{\epsilon}^F_{j,l}$, of the transmission essentially 
measures $\Delta F$, i.e. the difference between $F(x_1)$ and $F(x_2)$ 
with $|x_1-x_2| \simeq L/2^j$. If $\Delta F$ is small, which is generally 
true, the DWT is analogous to finite difference. The $F-\delta$ relation 
then yields  
%eq16 
\begin{equation}
\tilde{\epsilon}_{j,l} \simeq 
 -\frac{1}{F(l)_jaA[1+\delta(l)_j]^{a-1}}\tilde{\epsilon}^F_{j,l},
\end{equation} 
where $\tilde{\epsilon}_{j,l}$ is the WFC of the mass contrast 
$\delta$. $F(l)_j$ and $\delta(l)_j$ are, respectively, the mean 
flux and mean mass contrast in the spatial range 
$Ll/2^j \leq x < L(l+1)/2^j$. 

Using eq.(16), one can study the relation between $\tilde{\epsilon}_{j,l}$ 
and $\tilde{\epsilon}^F_{j,l}$ at each position $l$. At positions with
$|\delta(l)| <1$, i.e. clustering is weak, we have approximately
%eq17
\begin{equation}
\tilde{\epsilon}_{j,l} \simeq 
 -\frac{1}{aA}\frac{\tilde{\epsilon}^F_{j,l}}{F(l)_j}=
 -\frac{1}{aA}\frac{\tilde{\epsilon}^F_{j,l}}{e^{-A}}.
\end{equation}
At positions with $\delta(l)>1$, we have 
%eq18 
\begin{equation}
|\tilde{\epsilon}_{j,l}| \leq
\left  |\frac{1}{aA}\frac{\tilde{\epsilon}^F_{j,l}}{F(l)_j} \right |.
\end{equation}
For voids, i.e. $\delta(l) \simeq -1$, $\tilde{\epsilon}^F_{j,l}$
generally is zero, which has no contribution to $P_j$ and $R_j$. 
Thus, eqs.(17) and (18) give an upper limit to the DWT power spectrum 
of the mass field as
%eq19
\begin{equation}
P_j \leq \frac{1}{a^2A^2} P'_j,
\end{equation}
where
%eq20
\begin{equation}
P'_j=\frac{1}{2^j}\sum_{l=0}^{2^j-1}
  \left [\frac{\tilde{\epsilon}^F_{j,l}}{F(l)_j}.
   \right ]^2
\end{equation}

In the DWT analysis, $F(l)_j$ can be calculated by (Fang \& Feng 2000) 
%eq21
\begin{equation}
F(l)_j=\left(\frac{L}{2^j}\right)^{1/2} \epsilon^F_{j,l},
\end{equation}
where 
%eq22
\begin{equation}
\epsilon^F_{j,l} = \int F(x) \phi_{j,l}(x) dx.
\end{equation} 
$\phi_{j,l}(x)$ is the scaling function (Fang \& Thews 1998), 
and $\epsilon^F_{j,l}$ is called 
scaling function coefficient (SFC) of the transmission. With 
$\epsilon^F_{j,l}$, $P'_j$ can be calculated by
%eq23
\begin{equation}
P'_j = \frac{L}{2^{2j}}\sum_{l=0}^{2^j-1}
 \left[ \frac{\tilde{\epsilon}^F_{j,l}}{\epsilon^F_{j,l}}\right]^2.
\end{equation}

Fig. 1 plots $P^F_j$ and $P'_j$ for the transmitted flux of HS1700.  
Although there are differences between the power spectra $P'_j$ and
$P^F_j$, they have similar shape. On large scales, most places $l$ have
$|\delta(l)| <1$. Therefore, from eq.(17), the DWT power spectrum of mass
field on large scales is given by 
%eq24
\begin{equation} 
P_j \simeq\frac{1}{a^2A^2e^{-2A}}P^F_j. 
\end{equation} 
On small scales, more places
have $\delta(l) >1$.  Therefore, the DWT power spectrum of the mass field
is constrained by $P^F_j/a^2A^2e^{-2A}$ as a lower limit, and
$P'_j/a^2A^2$ as an upper limit. The number $e^{-A}$ is given by mean
transmission, and is in the range 0.5-0.9. Thus, the shape of $P_j$ is
well constrained by the shapes of $P^F_j$ and $P'_j$, i.e. flat on large
scales and rapidly decreasing on small scales.

Similarly, one can estimate the roughness $R_j$ of mass field by 
calculating $R^F_j$ and $R'_j$, where $R^F_j$ is defined by eq.(15), and
$R'_j$ is also given by eq.(15) but the WFC $\tilde{\epsilon}^F_{j,l}$ 
is replaced by with $\tilde{\epsilon}^F_{j,l}/\epsilon^F_{j,l}$. 
However, roughness is defined as a ratio of WFCs, and the correlation 
between $\tilde{\epsilon}^F_{j,l}$ and $\epsilon^F_{j,l}$ is not very
strong. For  roughness, the effect of replacing 
$\tilde{\epsilon}^F_{j,l}$ by $\tilde{\epsilon}^F_{j,l}/\epsilon^F_{j,l}$ 
is weak. Thus, roughness $R_j$ of mass field can directly be estimated
by $R^F_j$.

\section{Local power spectrum of hierarchical clustering}

The hierarchical clustering provides a skeleton for modeling the formation
of structures via the merging of dark halos. It is assumed
that the mass field formed by the non-linear evolution of the cosmic
gravitational clustering can be described by the linked-pair approximation,
or correlation hierarchy, i.e. the $n$-th irreducible correlation function
$\xi_n$ is given by the two-point correlation function $\xi_2$ as
$\xi_n = Q_n \xi_2^{n-1}$, where $Q_n$ is the hierarchical coefficient
(White 1979). It is, however, well known that correlation hierarchy with
constant coefficients $Q_3$ cannot match with the skewness of non-linear
mass field given by the perturbation calculation and N-body simulation
(e.g. Jing \& B\"orner 1998). Moreover, the local power spectrum of the
transmission (\S 3) strongly indicates that the underlying mass field of
a QSO's Ly$\alpha$ forest is intermittent. This implies that the
correlation hierarchy would be a problem on small scales. In this section,
we will find the applicable range of the correlation hierarchy by testing 
its predictions of the local power spectrum.

\subsection{Roughness of hierarchically clustered field}

The first test is on the roughness. Let us calculate the roughness $R_j$ 
with the linked pair approximation. For $n=4$, the hierarchical relation 
of mass field correlation functions is
%eq25
\begin{eqnarray}
\lefteqn{\langle \delta({\bf x_1}) \delta({\bf x_2}) \delta({\bf x_3})
\delta({\bf x_4})\rangle  =  } \\ \nonumber
 & &  Q^a_4[\langle \delta({\bf x_1}) \delta({\bf x_2})\rangle
\langle \delta({\bf x_2}) \delta({\bf x_3})\rangle
\langle \delta({\bf x_3}) \delta({\bf x_4})\rangle +
 {\rm cyc. \ 12 \ terms}]\\ \nonumber
  & & +Q^b_4[\langle \delta({\bf x_1}) \delta({\bf x_2})\rangle
\langle \delta({\bf x_1}) \delta({\bf x_3})\rangle
\langle \delta({\bf x_1}) \delta({\bf x_4})\rangle +
{\rm cyc. \ 4 \ terms}].
\end{eqnarray}
where $Q^a_4$ is for snake diagrams and $Q^b_4$ is for stars.

Because the samples of Ly$\alpha$ transmitted flux is 1-dimensional, to
calculate 1-D WFCs we use a projection of a 3-D distribution
$\delta({\bf x})$ onto 1-D as
%eq26
\begin{equation}
\tilde{\epsilon}_{j,l}=
\int_{-\infty}^{\infty} \delta({\bf x})\psi_{j,l}(x^1)
  \phi_{J',m}(x^2)\phi_{J',n}(x^3)dx^1dx^2dx^3,
\end{equation}
where $x^1$ is for the redshift direction, $x^2$ and $x^3$ are the
dimensions of the sky, and $\phi_{j,l}(x)$ is the scaling function (\S 3.4). 
The scaling function  $\phi_{j,l}(x)$ in eq.(26) plays the role of a
window function on scale $j$ at position $l$.
With eq.(26), a 1-D field with cross-section $L/2^{J'}\times L/2^{J'}$
along the $x^1$-direction can be decomposed as
%eq27
\begin{equation}
\delta({\bf x}) =
  \sum_{j}\sum_{l} \tilde{\epsilon}_{j,l}
   \psi_{j,l}(x^1)\phi_{J',m}(x^2)\phi_{J',n}(x^3).
\end{equation}
For a QSO Ly$\alpha$ absorption spectrum, $m$ and $n$ denote the position
of the QSO on the sky, and scale $J'$ is determined by the size of the
absorption clouds. Actually, we don't know the exact scale $J'$. However,
the details of $\phi_{J',m}(x^2)$ and $\phi_{J',n}(x^3)$ do not affect the
conclusions below.

Using the decomposition eq.(27), eq.(25) yields 
%eq28
\begin{eqnarray}
\lefteqn{\langle \tilde{\epsilon}_{j,l}^4\rangle_{ir}
   = Q^a_4B_a[\sum_{j',l'}\sum_{j'',l''}\sum_{j''',l'''}
     \sum_{j'''',l''''}
  \langle \tilde{\epsilon}_{j,l}\tilde{\epsilon}_{j',l'}\rangle
\langle \tilde{\epsilon}_{j'',l''}\tilde{\epsilon}_{j''',l'''} \rangle
\langle \tilde{\epsilon}_{j'''',l''''}\tilde{\epsilon}_{j,l}\rangle }
  \\ \nonumber
 & & \int\psi_{j,l}(x_2)\psi_{j',l'}(x_2)\psi_{j'',l''}(x_2)dx_2
\int\psi_{j,l}(x_3)\psi_{j''',l'''}(x_3)\psi_{j'''',l''''}(x_3)dx_3
  \\ \nonumber
& & + {\rm cyc. \ 12 \ terms}]  \\ \nonumber
  & & + Q^b_4B_b[\sum_{j',l'}\sum_{j'',l''}\sum_{j''',l'''}
\langle\tilde{\epsilon}_{j',l'}\tilde{\epsilon}_{j,l}\rangle
\langle\tilde{\epsilon}_{j'',l''}\tilde{\epsilon}_{j,l}\rangle
\langle\tilde{\epsilon}_{j''',l'''}\tilde{\epsilon}_{j,l}\rangle
    \\ \nonumber
 & & \int\psi_{j,l}(x_1)\psi_{j',l'}(x_1)\psi_{j'',l''}(x_1)
  \psi_{j''',l'''}(x_1)dx_1  + {\rm cyc. \ 4 \ terms}],
\end{eqnarray}
where subscript $ir$ stands for the irreducible correlation function, 
$B_a=\int \phi^3_{J',m}(x^2)\phi^3_{J',n}(x^3)dx^2dx^3$
and $B_b=\int \phi^4_{J',m}(x^2)\phi^4_{J',n}(x^3)dx^2dx^3$. These
constants can be absorbed into coefficients $Q^a_4$ and $Q^b_4$,
respectively. We will no longer show the two constants
explicitly.

Because, the covariance
$\langle \tilde{\epsilon}_{j,l}\tilde{\epsilon}_{j',l'}\rangle$
is generally quasi $j$-diagonal at least for the clustering referred
to QSO Ly$\alpha$ absorption (Feng \& Fang 2000), the r.h.s. of eq.(28)
is dominated by the terms
$\langle \tilde{\epsilon}_{j,l}^2 \rangle^{3}$. We have then
%eq29
\begin{equation}
\langle \tilde{\epsilon}_{j,l}^4 \rangle_{ir} \simeq
[12 Q_4^a (a^3_j)^2 + 4Q^b_4 a^4_j]
  \langle \tilde{\epsilon}_{j,l}^2 \rangle^{3}.
\end{equation}
where the factors $a^3_j$ and  $a^4_j$ are given by
%eq30
\begin{equation}
a^3_j=\int \psi_{j,l}^3(x)dx=
\left(\frac{2^j}{L}\right)^{1/2} \int \psi^{3}(x)dx,
\end{equation}
%eq31
\begin{equation}
a^4_j=\int \psi_{j,l}^4(x)dx=
 \frac{2^j}{L} \int \psi^{4}(x)dx,
\end{equation}
where $\psi(x)$ is the basic wavelet (Fang \& Thews 1998).

Thus, from eqs.(29) and (15), which include only the irreducible
correlation, the roughness of a hierarchically clustered field is given by
%eq32
\begin{equation}
R_j= Q^R_4 2^j P_j,
\end{equation}
where the hierarchical coefficient $Q^{R}_4$ is given by
%eq33
\begin{equation}
Q^{R}_4 = A_aQ_4^a + A_b Q_4^b,
\end{equation}
and
%eq34
\begin{equation}
A_a= \frac{4}{L} \left[\int \psi^{3}(x)dx \right]^2
\end{equation}
%eq35
\begin{equation}
A_b= \frac{4}{3}\frac{1}{L} \int \psi^{4}(x)dx.
\end{equation}
As expected, in the linked-pair approximation the roughness is
completely determined by the power spectrum $P_j$ (two-point correlation
function) and coefficient  $Q_4$.

The dotted-line curve in Fig. 4 is $R_j$ calculated by eq.(32), 
but $P_j$ is replaced by $P_j'$. Therefore, the dotted-line curve 
in Fig. 4 actually is an upper limit to the linked-pair predicted 
roughness for mass field. Fig. 4 shows that, with a constant $Q^R_4$ 
fitting, the linked-pair predicted $R_j$ cannot match the observation 
on scales less than $0.3$ h$^{-1}$ Mpc, as the predicted $R_j$ is 
smaller on smaller scales, while the observed result is larger on smaller 
scales. This result depends only on the shape of the power spectrum.
The decrease of the linked-pair predicted roughness $R_j$ 
with $j$ on scales less than $0.3$ h$^{-1}$ Mpc is due to the decrease 
of $P'_j$ and $P^F_j$ with $j$. Since the shape of mass field power 
spectrum $P_j$ is constrained by $P'_j$ or $P^F_j$, one can conclude that 
the hierarchical clustering model with constant $Q_4$ is a good 
approximation on scales $ > 0.2$ h$^{-1}$ Mpc. 

\subsection{Scale-scale correlations of hierarchical clustered
    field}

Figs. 2 and 3 show that some spikes on different scales $j$ have the
same physical position. That is, the locations of spikes on different
scales are correlated.  For instance, a singular structure like
$\rho (r)\propto r^{-\alpha}$ (\S 1) is a place where the large
density difference ($|\rho(x+r)-\rho(x)|$) events on different scales
$r$ are in phase. This is known as {\it scale-scale correlation}. It can be
measured by the correlations between $P_{j,l}$ and $P_{j',l'}$ with
$j\neq j'$. In the hierarchical clustering model, the scale-scale
correlation is also determined by the linked-pair approximation. This
provides the second test for the hierarchical clustering model.

To measure this correlation, we use the normalized local power
spectrum as
%eq36
\begin{equation}
p_{j,l}=\frac{P_{j,l}}{P_j}.
\end{equation}
Obviously, $\langle p_{j,l}\rangle =1$. The correlation
between local power spectra on scales $j$ and $j+1$ can be calculated by
%eq37
\begin{equation}
C^{(2)}_j= \frac{1}{2^{j+1}}\sum_{l=0}^{2^{(j+1)}-1} p_{j,[l/2]} p_{j+1,l}
\end{equation}
where the brackets denote the integer part of the quantity enclosed.
$C_j^{(2)}$ is greater than 1 if the spikes on the scale $j$ have higher
than random probability of appearing at the same physical position as
$j+1$ spikes.
It should be emphasized that although $C^{(2)}_j$ is also 4-th order
statistics, and it is independent of $R_j$. Correlation (37) can be generalized
to any pair of scales $j$ and $j'$.

Using a similar approach for deriving eq.(32), one can find the expression of
$C^{(2)}_j$ for a hierarchically clustered field. However, it is rather
complicated. Noting that power on small $j$ is always larger than that on
large $j$, especially on small scales (large $j$), we have
%eq38
\begin{equation}
C^{(2)}_j \simeq Q^{C}_4 2^j P_j,
\end{equation}
where the hierarchical coefficient $Q^C_4$ is
%eq39
\begin{equation}
Q^{C}_4=A'_aQ_4^a + A'_bQ_4^b,
\end{equation}
and
%eq40
\begin{equation}
A'_a = \frac{16}{L}\left[\int\psi^{2}(x)\psi(2x)dx\right]^2
       +  \frac{4}{L}\left[\int\psi^{3}(x)dx\right]^2
\end{equation}
%eq41
\begin{equation}
A'_b=\frac{2}{L}\int \psi^{2}(x)\psi^2(2x)dx.
\end{equation}
Similar to eq.(32), in the linked-pair approximation the scale-scale 
correlations are completely determined by the power spectrum $P_j$ 
(two-point correlation function) and coefficient  $Q_4$.

Again, to test the lined-pair prediction (38), we should calculate 
$C^{(2)}_j$ of mass field. Since $C^{(2)}_j$ is also defined by a 
ratio of WFCs, the effect of $F-\delta$ nonlinearity on $C^{(2)}_j$ 
is also weak. We calculate the correlation $C^{(2)}_j$ by eq.(37) for the 
sample HS1700+64. The result is plotted in Fig. 5. It shows a strong 
correlation $C^{(2)}_j$ on small scales ($j >9$), and in this range 
we have approximately a power law as
%eq42
\begin{equation}
 C^{(2)}_j \propto 2^{j\mu},
\end{equation}
with index $\mu \simeq 0.9$. 

Fig. 5 also shows the linked-pair predicted $C^{(2)}_j$ given by 
eq.(38), where $P_j$ is replaced by $P'_j$. The hierarchical clustering 
with a constant value of $Q_4$ basically are able to fit the observation on 
scales $j \leq 9$ or $ > 0.2$ h$^{-1}$ Mpc, but fail on small scales. 
Similar to Fig. 4, this result depends only on the shape of the power 
spectrum. The over-all normalization of the power spectrum can be absorbed
in the factor $Q_4^C$.

\subsection{Scale-dependent $Q_4$}

To apply the hierarchical clustering model to small scales, one can assume 
that the coefficients 
$Q^R_4$ and $Q^C_4$ are scale dependent. That is, the deviation between
the linked-pair prediction and the observation shown in Figs. 4 and 5 could
be eliminated by using scale-dependent $Q^R_4$ and $Q^C_4$ given by
%eq43
\begin{equation}
Q^R_4=\frac {R_j}{2^j P_j},
\end{equation}
and
%eq44
\begin{equation}
Q^C_4= \frac{C^{(2)}_j}{2^j P_j}.
\end{equation}

However, eqs.(43) and (44) pose a new problem. In the linked-pair
approximation eq.(16), there are two coefficients $Q_4^a$ and $Q_4^b$.
Therefore, the 4th order correlation hierarchy, $\xi_4=Q_4 \xi_2^3$,
with scale-dependent $Q_4$ is reasonable only if the two coefficients
$Q_4^a$ and $Q_4^b$ possess the same scale-dependence. In other words,
that $Q^R_4$ and $Q^C_4$ have the same scale-dependence is a necessary
condition to use the 4th order correlation hierarchy. Therefore, 
whether $Q_4^R$ and $Q_4^C$ have the same scale-dependence is a 
test of the hierarchical clustering with scale-dependent $Q$.

This test, however, needs the DWT power spectrum of mass field. An 
upper limit spectrum given by $P'_j$ seems not to be enough. 
Yet, our purpose is not to obtain the values of $Q^R_4$ and $Q^C_4$, but 
their behavior of $j$-dependence. Fig. 6 shows the $j$-dependencies of 
$Q^R_4$ and $Q^C_4$ calculated by eqs.(43) and (44) with $P^F_j$(lower panel) 
and $P'_j$(upper panel). Although the values of $Q^R_4$ and $Q^C_4$ given 
by $P_j$ and $P'_j$ are significantly different from each other on small 
scales, the two hierarchical coefficients 
($Q^R_4$, $Q^C_4$), and so ($Q_4^a$, $Q_4^b$)
have the same $j$ dependent behavior in the range from $j=6$ to 13.
Since  $P_j$ has the similar shape as $P^F_j$ and $P'_j$, Fig. 6 strongly 
indicates that one scale-dependent $Q_4$ probably is reasonable in the 
scale range from $2.5$ h$^{-1}$ Mpc to few tens h$^{-1}$ kpc.

\section{Conclusions and discussions}
 
The local DWT power spectrum provides an unified description of the
power distribution of clustering in both scale space and
physical space. That is, it contains information about the amplitudes
and phases of the density perturbations. With this tool, the
central part (or the variance) of the probability distribution function
(PDF) of the density difference $|\rho(x+r)-\rho(x)|$ can be measured
by the averaged local DWT power spectrum, while the tail of the PDF
can be measured by the roughness of the local DWT power spectrum.
Therefore, it is effective to describe the non-linear evolution of
the cosmic mass field.

With the local power spectrum of Ly$\alpha$ transmitted flux of QSO
HS1700, we find that the underlying cosmic mass field of the transmitted
flux with redshift around $z \simeq 2.2$ can be described by the 
hierarchical clustering model on the physical scales from 2.5 h$^{-1}$ 
Mpc to few tens h$^{-1}$ kpc. But the non-linear features of clustering 
are different on different scale ranges. 1. On physical scales larger 
than $\sim 1.3$ h$^{-1}$ Mpc, the field is almost Gaussian. 2. On scales 
1.3 h$^{-1}$ Mpc - $0.3$ h$^{-1}$ Mpc, the field is consistent with the 
correlation hierarchy with a constant value for the coefficient $Q_4$. 3. 
On scales less than 300 h$^{-1}$ kpc, the field is no longer Gaussian, 
but essentially intermittent. In this case, the field can 
still be fitted 
by the correlation hierarchy, but the coefficient, $Q_4$, should be 
scale-dependent. The above three points are strongly
supported by the following result: the scale dependencies of $Q_4$
given by two statistically independent measures, i.e. $Q_4^R$ by 
the roughness and $Q_4^C$ by scale-scale correlation are the same 
in the entire scale range considered.

In the clustering of cosmic mass field, the mass perturbation on small
scales evolved into the non-linear regime early, followed by the large scale
perturbations. Therefore, the above-mentioned result implies that
the non-linear evolution of the cosmic mass field underwent the following
stages: Gaussian -- hierarchical clustering with constant $Q$ -- 
intermittent or hierarchical clustering with scale-dependent $Q$.

The local power spectrum can detect the scale range of the
non-linear evolution as well as the intermittent behavior of the cosmic
mass field. It is, hence, a potential discriminator amongst models
of structure formation. An advantage
of the intermittent discriminator is that the non-linear behavior of
the cosmic mass field can be tested not only with high density
objects, like the centers of massive halos, but also with low
density objects, like QSO's Ly$\alpha$ forests.

It is not difficult to generalize this analysis to $Q_n$ with $n>4$ if
better data are available. Generally, the $n$th correlation hierarchy
$\xi_n = Q_n\xi_2^{n-1}$ can be tested by calculating the $n$th moment of
the PDF of the WFCs.

\acknowledgments

We thank Dr. Wolung Lee for his support to this work. We also thank
Dr. D. Tytler for kindly providing the data of the Keck spectrum
HS1700+64. HZ acknowledges the support from Arizona State University through
a Chair's Summer Research Fellowship. PJ would like to thank Dr Jill 
Bechtold and Jennifer Scott for
useful discussions. PJ would also like to thank Dr. Robert Maier for 
helpful discussions.

\clearpage

%\epsscale{0.8} 
\plotone{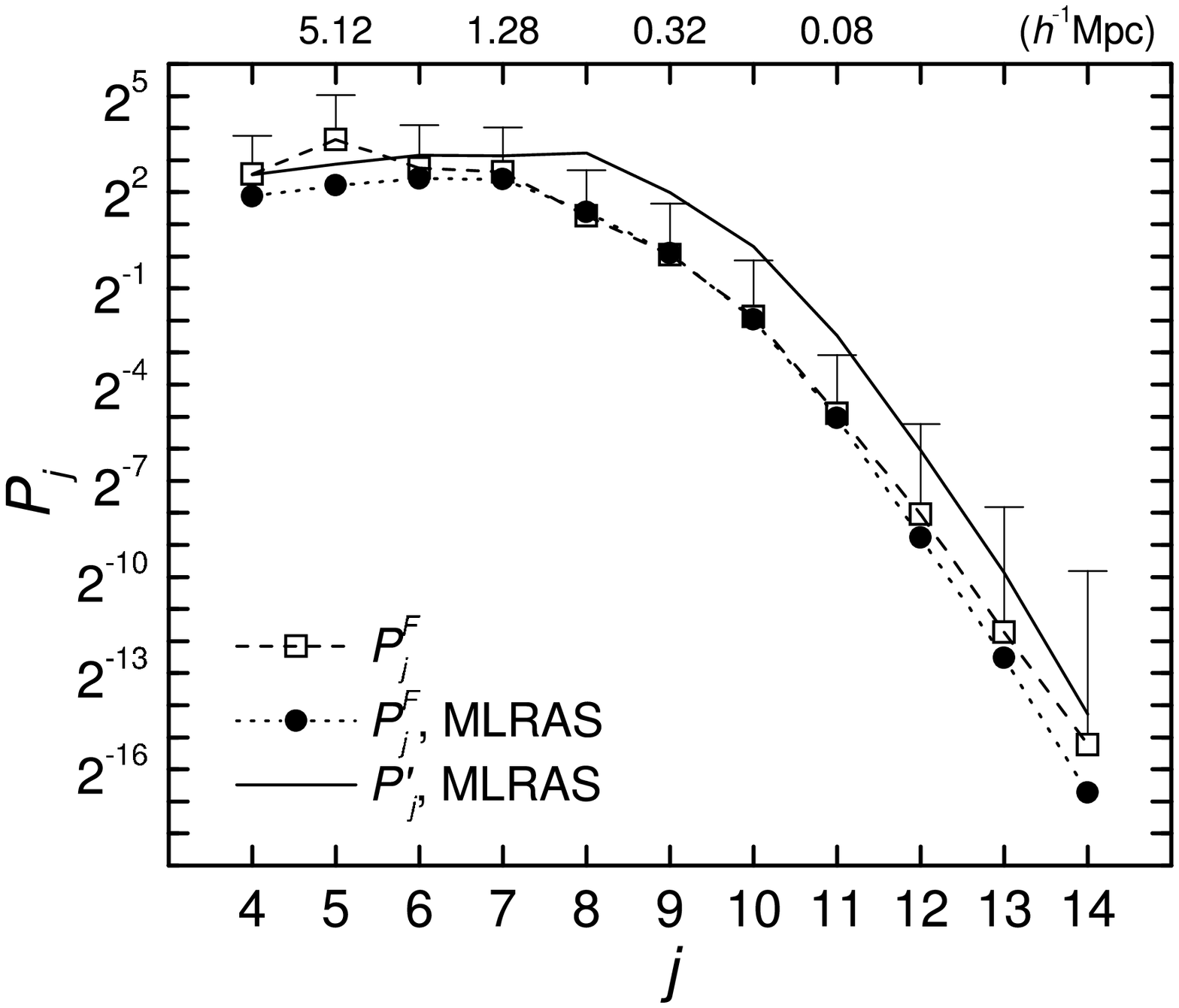} 
\figcaption[f1.eps]{The DWT power spectrum $P^F_j$ of the original transmitted 
flux of HS1700+64, the metal-line-removed-and-smoothed (MLRAS) flux, and 
$P'_j$ as defined by eq.(37). The horizontal
axis represents the physical scale of $2^{15-j}\times 5 h^{-1}$ kpc. The
values of $P^F_j+\sigma^{p}_{j}$ are also shown by error bars. 
\label{fig1}}

\clearpage
%\epsscale{0.8}
\plotone{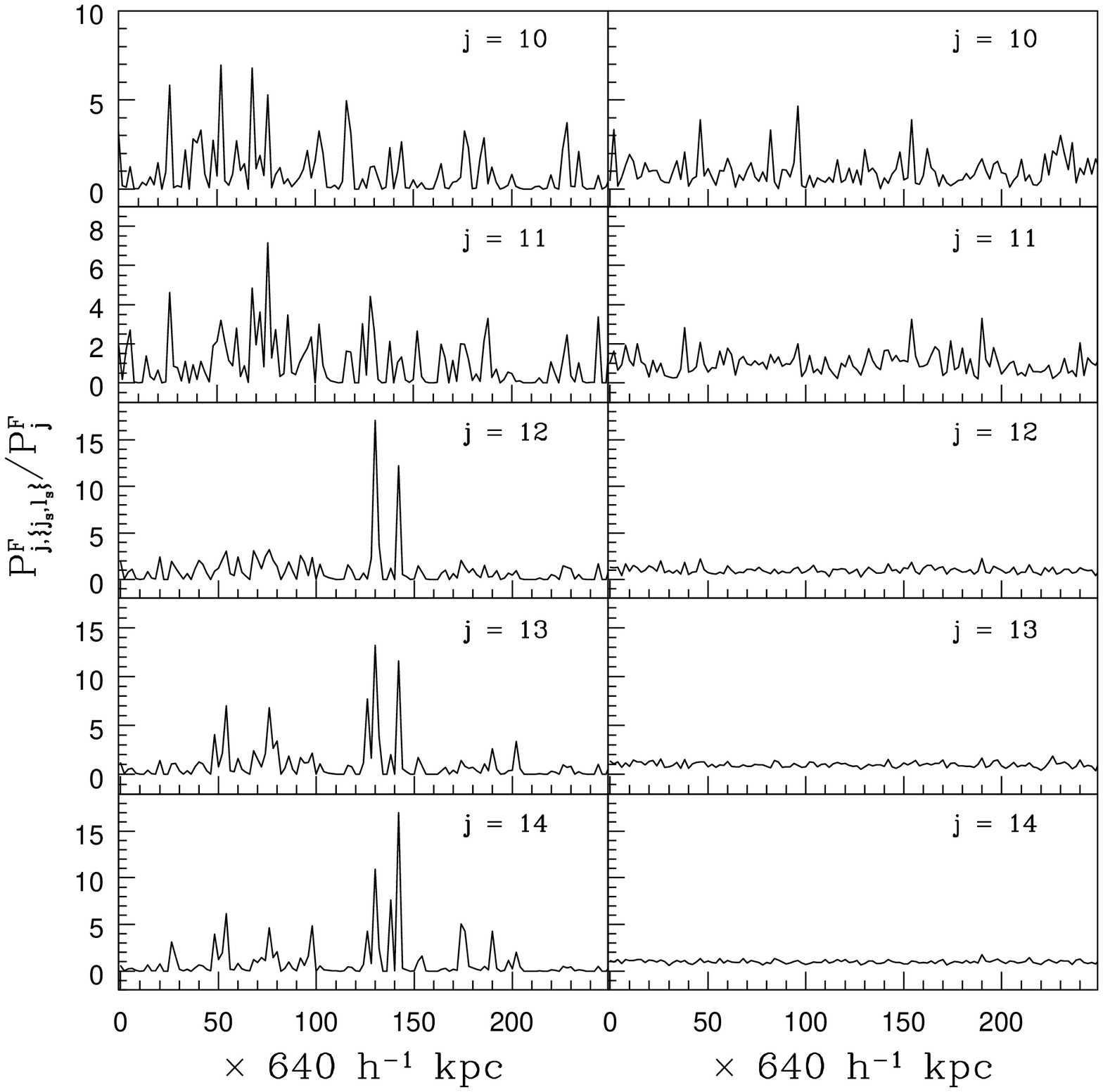}
\figcaption[f2.eps]{The normalized local power spectrum $P^F_{j,l}$/$P^F_j$
($j=10 - 14$) of the transmitted flux of HS1700+64 (left panels),
and its phase-randomized (PR) counterparts (right panels).
The horizontal axis represents relative proper distance 
($\approx l \times 5 \times 2^{15-j}$ h$^{-1}$ Kpc) in an 
Einstein-de Sitter universe.
\label{fig2}}

\clearpage
%\epsscale{0.8}
\plotone{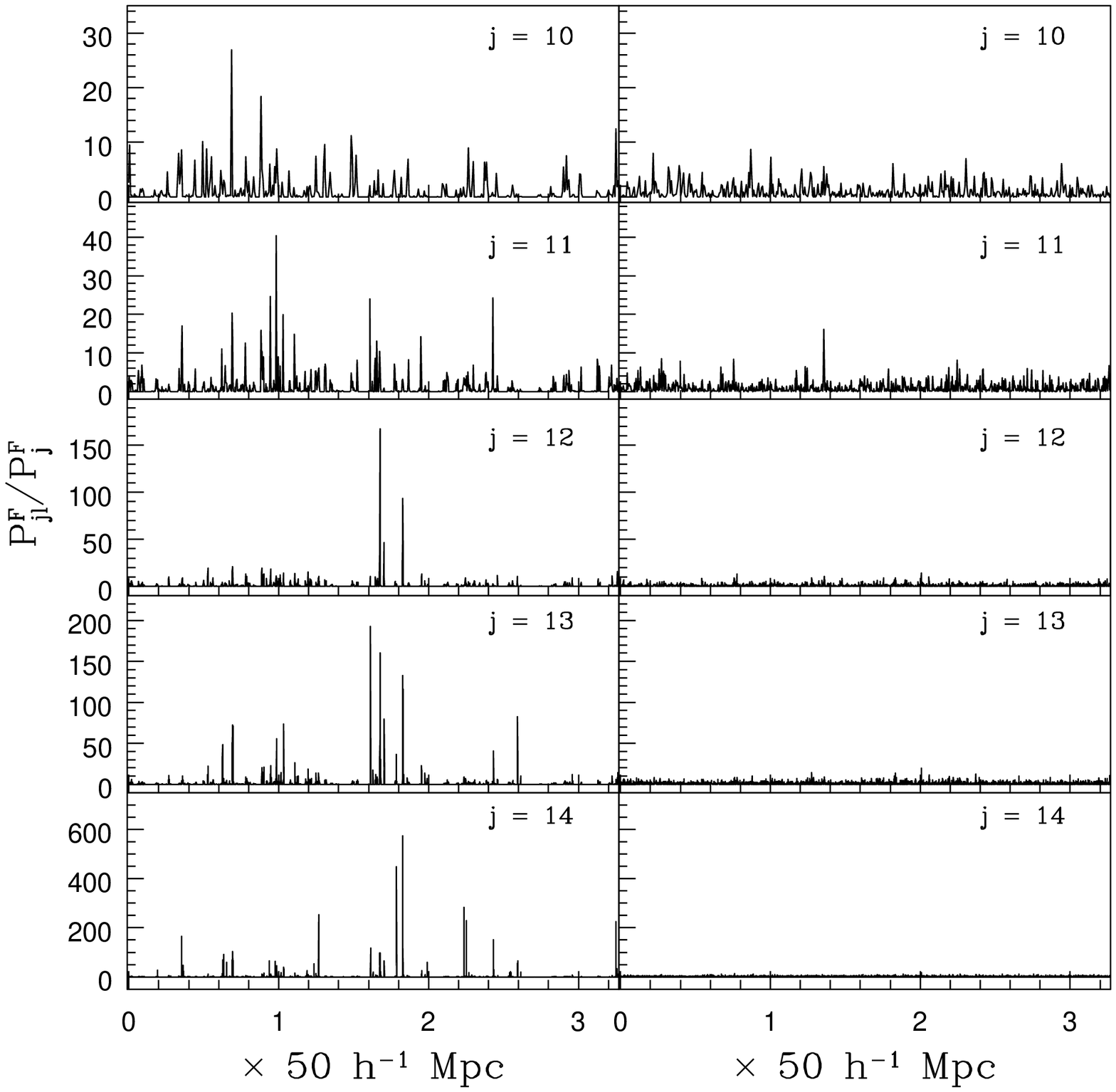}
\figcaption[f3.eps]{The normalized local power spectrum,
$P^F_{j,\{j_s,l_s\}}/P^F_j$, of the transmitted
flux of HS1700+64 (left panels), and its phase-randomized (PR) counterpart
(right panels) with $j_s = 8$. The horizontal axis represents relative
proper distance ($\approx l_s \times 640$ h$^{-1}$ Kpc) in an 
Einstein-de Sitter universe. 
\label{fig3}}

\clearpage
%\epsscale{0.7}
\plotone{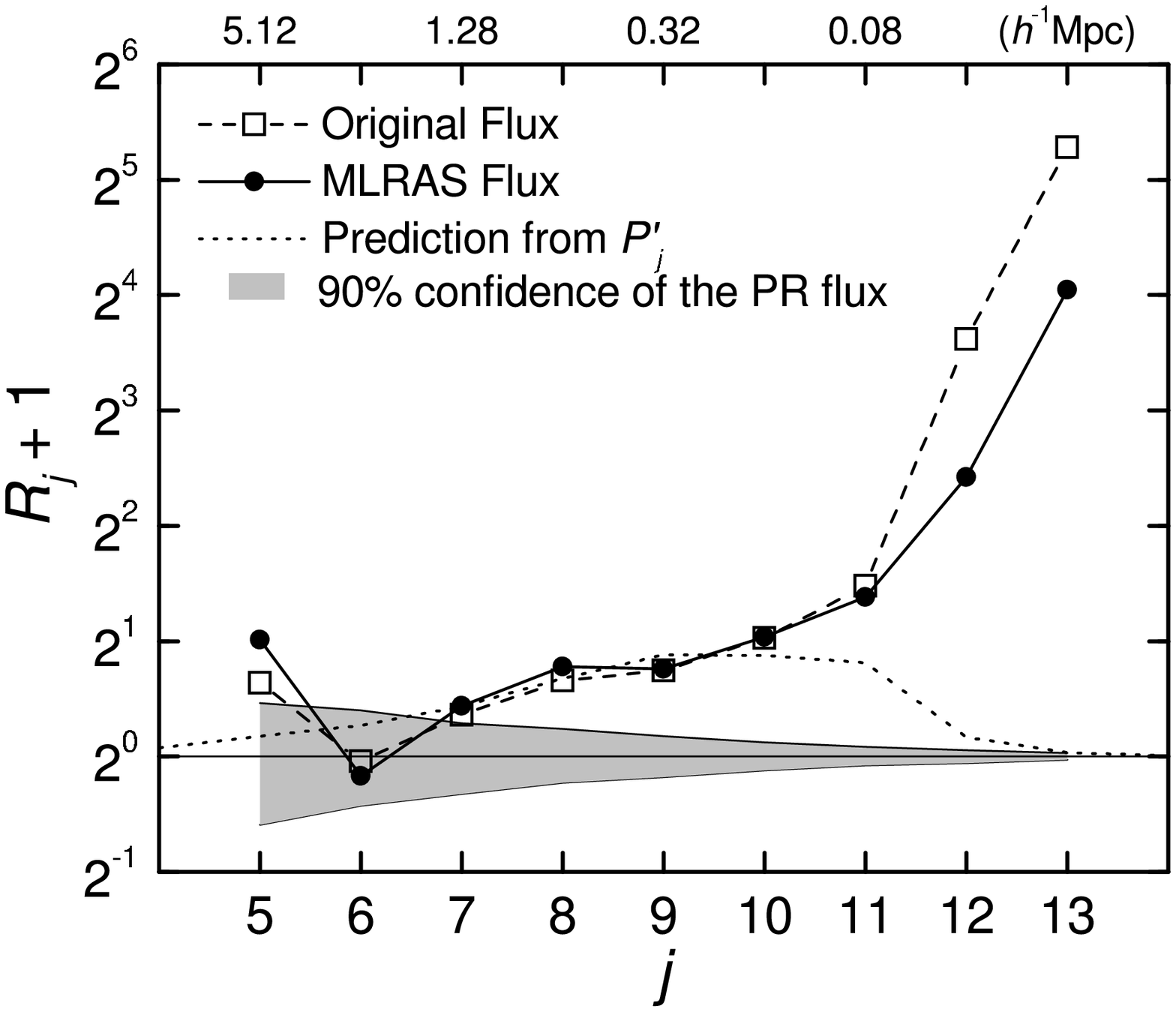}
\figcaption[f4.eps]{The roughness, $R^F_j$, of the original flux, the
metal-line-removed-and-smoothed (MLRAS) flux, and the prediction from the
correlation hierarchy model with a constant $Q_4^R$. The gray band is 
the 90\% confidence of the phase-randomized (PR) flux. The vertical axis is
$R_j+1$. This made easy for the logarithmic scale of $R_j$ in the range 
$0>R_j >-1$.
\label{fig4}}

\clearpage
%\epsscale{0.8}
\plotone{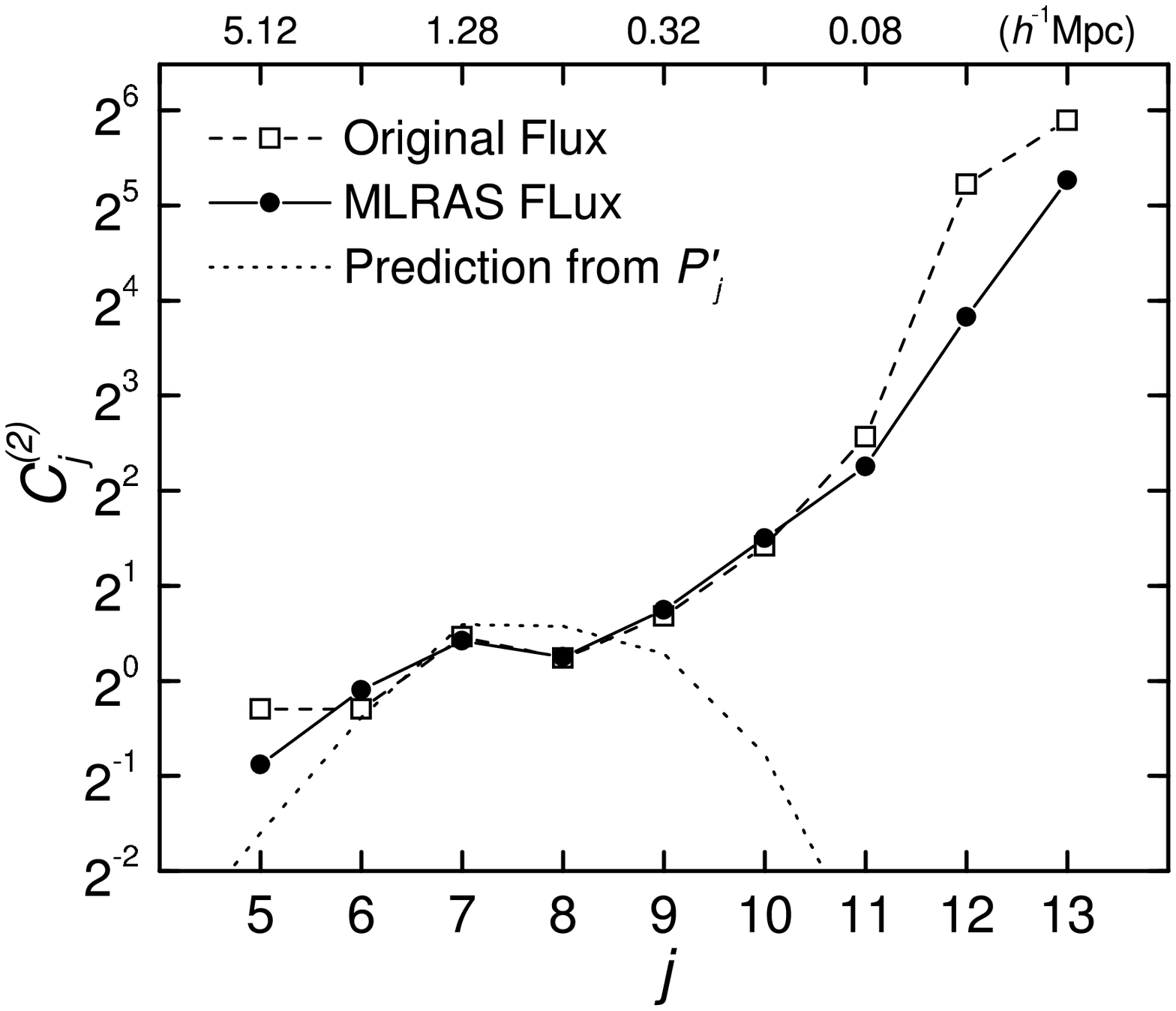}
\figcaption[f5.eps]{The correlation, $C^{(2)}_j$, of the original flux,
the metal-line-removed-and-smoothed (MLRAS) flux, and the prediction from the
correlation hierarchy model with a constant $Q_4^C$.
\label{fig5}}

\clearpage
\epsscale{0.85}
\plotone{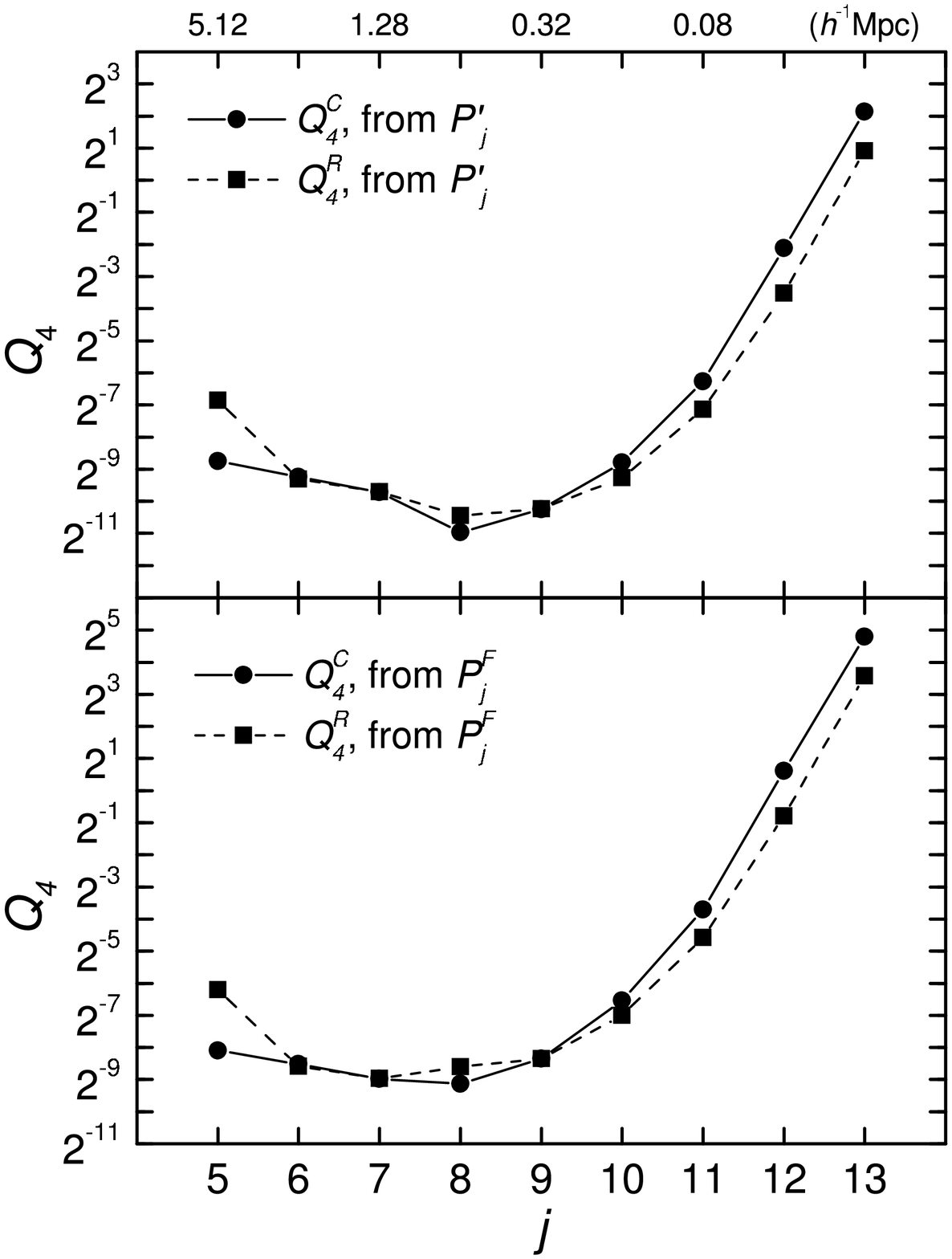}
\figcaption[f6.eps]{The scale-dependence of the hierarchical
coefficients $Q^C_4$ and $Q^R_4$ given by eqs.(42) and (43)
with power spectrum $P^F_j$ (lower panel) and $P'_j$ (upper 
panel.)
\label{fig6}}

\end{document}